\documentclass[preprint,12pt,authoryear]{elsarticle}

\usepackage{amsmath, amsfonts, amssymb, amsthm}
\usepackage{hyperref}
\usepackage{color}
\usepackage{graphicx}
\usepackage{datetime}
\usepackage{dsfont}
\usepackage{subcaption}
\usepackage{array}
\usepackage{multirow}

\usepackage{url}

\theoremstyle{plain}

\theoremstyle{definition}

\theoremstyle{remark}

\journal{Finance Research Letters}

\allowdisplaybreaks

\begin{document}

\begin{frontmatter}

\title{Portfolio Optimization with Feedback Strategies Based on Artificial Neural Networks}

\author[YK]{Yaacov Kopeliovich}
\author[MP]{Michael Pokojovy\corref{cor}}

\cortext[cor]{Correspondence: \url{mpokojovy@odu.edu}}

\affiliation[YK]{organization={Finance Department, University of Connecticut},
            addressline={2100 Hillside Road}, 
            city={Storrs},
            postcode={06269}, 
            state={CT},
            country={USA},
            orcid={, ORCID:0000-0001-7729-1684}}

\affiliation[MP]{organization={Department of Mathematics and Statistics, Old Dominion University},
            addressline={4300 Elkhorn Ave}, 
            city={Norfolk},
            postcode={23529}, 
            state={CT},
            country={USA},
            orcid={, ORCID:0000-0002-2122-2572}}

\begin{abstract}
    With the recent advancements in machine learning (ML), artificial neural networks (ANN) are starting to play an increasingly important role in quantitative finance. Dynamic portfolio optimization is among many problems that have significantly benefited from a wider adoption of deep learning (DL). While most existing research has primarily focused on how DL can alleviate the curse of dimensionality when solving the Hamilton-Jacobi-Bellman (HJB) equation, some very recent developments propose to forego derivation and solution of HJB in favor of empirical utility maximization over dynamic allocation strategies expressed through ANN. In addition to being simple and transparent, this approach is universally applicable, as it is essentially agnostic about market dynamics. To showcase the method, we apply it to optimal portfolio allocation between a cash account and the S\&P 500 index modeled using geometric Brownian motion or the Heston model. In both cases, the results are demonstrated to be on par with those under the theoretical optimal weights assuming isoelastic utility and real-time rebalancing. A set of~\texttt{R} codes for a broad class of stochastic volatility models are provided as a supplement.
    {\it Published paper:} \url{https://doi.org/10.1016/j.frl.2024.106185}
\end{abstract}


\begin{highlights}
    \item A deep learning framework for optimal asset allocation under stochastic volatility via empirical utility maximization is presented.
    \item The approach is validated over a variety of scenarios based on two market models calibrated on S\&P 500 and VIX data.
    \item The developments can serve as a blueprint for how empirical utility maximization can be applied to virtually any type of asset allocation problems. 
\end{highlights}

\begin{keyword}


    Asset allocation \sep deep learning \sep empirical risk minimization \sep stochastic volatility \sep Heston model.

    \JEL G11 \sep C61
\end{keyword}

\end{frontmatter}

\section{Introduction}

\cite{Ma1952} published his paper on asset allocation in the presence of risk and revolutionized the financial world. The ideas expressed in that paper led to the emergence of modern investing strategies of passive investing, gave rise to the capital asset pricing (CAPM) model and revolutionized our thinking in terms of risk management and investing into efficient portfolios. However, \cite{Ma1952} paper considered a problem, in which investors allocate their assets statically instead of dynamically trading them over a given investment period. The latter problem of optimal allocation was solved by~\cite{Mer1969}. In his paper, the optimal allocation problem was viewed through the lens of utility theory where the investor seeks to maximize a certain utility function through a proper choice of the trading strategy. As a result, portfolio  allocation problem could be formulated as a stochastic control problem. Deriving and solving the associated Hamilton-Jacobi-Bellman (HJB) partial differential equation (PDE), \cite{Mer1969} was able to obtain the strategy for optimal allocation. In fact, \cite{Ma1952} problem was then reduced to a class of quadratic utilities. \cite{Mer1969} paper caused a second revolution, mostly in academe, and led to hundreds of papers, where optimal allocation problem were examined under different utilities and different assumptions of stochastic asset allocation behavior. 

In parallel, another technological revolution was brewing in the computer science world spanning decades of work, from the emergence of first machine learning (ML) models, to most recent breakthroughs in transformers and large language models giving rise to human-like bots like ChatGPT. It is, therefore, a natural question whether some of the emerging AI techniques can be used to solve contemporary problems in financial management, for example asset allocation. In this paper, we demonstrate that such synergy is indeed possible and can lead to new solution approaches for Merton-style dynamic asset allocation. In other words, we consider the Merton problem and attempt to apply AI tools (specifically, artificial neural networks (ANN)) to determine the optimal trading strategy (allocation) using deep learning (DL) methods. 

Our approach to optimal asset allocation strategy is based on recent advancements in DL, which enables us to represent trading strategies as ANN feedback control functions and train the latter using stochastic gradient descent (SGD). Unlike most of existing works that use DL~\citep{BaHuLaPha2022, HaJeE2018} for solving the HJB equation that arise in stochastic control problems akin to that considered by~\cite{Mer1969}, our strategy allows to entirely forego the derivation and solution of HJB in favor of empirical utility maximization~\citep{HaE2016, ReSo2023} with respect to ANN weights. In addition to simplicity, flexibility and computational efficacy, this approach is universally applicable as it is agnostic about the market dynamics that can be supplemented as a ``black box.'' A central question we aimed to investigate is whether this new DL-based approach to asset allocation problems is practical. For example, a similar attempt was made by~\cite{vanStaFoLi2023}, but the implementation involved running 2.56 million Monte Carlo simulations of the market dynamics that may take days to obtain a solution for a simple problem and, thus, may be somewhat impractical compared to traditional schemes based on analytical solutions or numerical approximations. Our goal was also to make the approach be as modular as possible and not depend on the underlying model or be tied to a specific HJB PDE. We believe to have accomplished both tasks and present our findings in this paper.

The structure of the paper is as follows. In Section~\ref{SECTION:STOCHASTIC_VOLATILITY_FRAMEWORK}, we formally present the Merton optimal portfolio problem. An asset following~\cite{He1993} stochastic volatility dynamics was adopted to highlight our approach is sufficiently robust to handle complicated dynamics of underlying assets. In Section~\ref{SECTION:FEEDBACK_CONTROL}, we describe how ANN feedback control can be trained to solve the Merton's problem. Real-world data application examples are presented in Section~\ref{SECTION:EXAMPLES}. Summary and conclusions are presented in Section~\ref{SECTION:SUMMARY_AND_CONCLUSIONS}. Euler-Maruyama discretization is finally given in~\ref{SECTION:DISCRETIZATION_APPENDIX}.

\section{Merton Optimal Portfolio Problem}
\label{SECTION:STOCHASTIC_VOLATILITY_FRAMEWORK}

Let $P_{t}$ be the price of a riskless asset continuously compounded at a constant rate $r$ and $S_{t}$ denote the price of a risky asset with a constant drift $\mu$ and stochastic volatility $\sqrt{Y_{t}}$. Under~\cite{He1993} stochastic volatility model, the market dynamics reads as
\begin{align}
    \label{EQUATION:MARKET_MODEL_1}
    \mathrm{d}P_{t} &= r P_{t}, & P_{t_{0}} &= p > 0, \\
    \label{EQUATION:MARKET_MODEL_2}
    \mathrm{d}S_{t} &= \mu S_{t} \mathrm{d}t + \sqrt{Y_{t}} \, S_{t} \mathrm{d}B^{S}_{t}, & S_{t_{0}} &= s > 0, \\
    \label{EQUATION:MARKET_MODEL_3}
    \mathrm{d}Y_{t} &= \kappa(\theta - Y_{t}) \mathrm{d}t + \sigma_{Y} \sqrt{Y_{t}} \, \mathrm{d}B^{Y}_{t}, &
    Y_{t_{0}} &= y > 0
\end{align}
with a correlated bivariate Wiener process $(B^{S}_{t}, B^{Y}_{t})_{t \geq 0}$ defined on a filtered probability space $(\Omega, (\mathcal{F}_{t})_{t \in [0, T]}, \mathbb{P})$ such that
\begin{equation}
    \label{EQUATION:NOISE_CORRELATION}
    \operatorname{Var}[\mathrm{d}B^{S}_{t}] = \mathrm{d}t, \quad
    \operatorname{Var}[\mathrm{d}B^{Y}_{t}] = \mathrm{d}t, \quad
    \operatorname{Cov}[\mathrm{d}B^{S}_{t}, \mathrm{d}B^{Y}_{t}] = \rho \, \mathrm{d}t
\end{equation}
for some $-1 \leq \rho \leq 1$. Here, $\theta > 0$ is the long-term mean, $\kappa > 0$ is the mean reversion speed and $\sigma_{Y} > 0$ is a constant ``volatility of volatility.'' Under the Feller condition $2 \kappa \theta > \sigma_{Y}^{2}$, Equations~\eqref{EQUATION:MARKET_MODEL_1}--\eqref{EQUATION:MARKET_MODEL_3} admit a unique strong solution $(P_{t}, S_{t}, Y_{t})$ such that $Y_{t} > 0$ holds $\mathbb{P}$-a.s.~for $t \geq t_{0} \geq 0$.

Letting $\pi_{t}$ denote the fraction of wealth invested in the risky asset, the wealth dynamics is then given by
\begin{align}
    \label{EQUATION:WEALTH_EVOLUTION}
    \mathrm{d}W_{t} &= W_{t} \big((1 - \pi_{t}) \, \mathrm{d} P_{t}/P_{t} + \pi_{t} \, \mathrm{d} S_{t}/S_{t}\big), & W_{t_{0}} = w
\end{align}
where $w > 0$ is the initial wealth. To emphasize the dependence of $W$ on $\pi$ we will sometimes write $W^{\pi}_{t}$.

Selecting a finite time horizon $T$, in the absense of intermediate consumption, the expected utility functional is assumed to have the form
\begin{equation}
    \label{EQUATION:UTILITY_FUNCTIONAL}
    J(t, p, s, y, w; \pi) = \mathbb{E}\big[u(W^{\pi}_{T}) \,\big|\, P_{t} = p, S_{t} = s, Y_{t} = y, W_{t}^{\pi} = w\big]
\end{equation}
where $u \colon (0, \infty) \to \mathbb{R}$ is a utility function selected to be the isoelastic utility
\begin{equation}
    \label{EQUATION:ISOELASTIC_UTILITY}
    u(w) = \begin{cases}
        \frac{w^{1 - \eta} - 1}{1 - \eta}, & \eta \neq 1 \\
        \ln(w), & \eta = 1
    \end{cases} \quad \text{ for } \eta \geq 0.
\end{equation}
The objective $J(\cdot)$ is then maximized over a suitably defined set $\mathcal{A}_{t, \boldsymbol{x}}$ of self-financing admissible control processes associated with the initial values $t$ and $\boldsymbol{x} = (p, s, y, w)$:
\begin{equation}
    \label{EQUATION:UTILITY_MAXIMIZATION_PROBLEM}
    J(t, p, s, y, w; \pi) \to \max_{\pi \in \mathcal{A}_{t, \boldsymbol{x}}}.
\end{equation}
No borrowing or short-selling constraints are imposed in this paper, i.e., the stock weight $-\infty \leq \pi_{t} \leq \infty$ can be arbitrary.

Under appropriate conditions, optimization over admissible controls $\pi_{t} \in \mathcal{A}_{t, \boldsymbol{x}}$ can be reduced to optimizing over Markov control policies~\citep{FleSo2006}, which allow to express the stock weight $\pi_{t}$ in the feedback form
\begin{equation}
    \label{EQUATION:FEEDBACK_CONTROL}
    \pi_{t} = \pi_{t}(\boldsymbol{x}).
\end{equation}
According to~\cite{Za2001}, the optimal weight can be computed as
\begin{equation}
    \label{EQUATION:OPTIMAL_CONTROL_ZARIPHOPOLOU}
    \pi^{\ast}(t, w, y) = \mathop{\operatorname{arg\,max}}_{\pi \in \mathbb{R}}
    \Big(\tfrac{1}{2} y \pi^{2} w^{2} V_{ww} + \rho \pi w y V_{wy} + (\mu - r) \pi w V_{w}\Big)
\end{equation}
where the value function 
\begin{equation}
    \label{EQUATION:VALUE_FUNCTION}
    V(t, w, y) = \max_{\pi \in \mathcal{A}_{t, \boldsymbol{x}}} J(t, w, y; \pi)
\end{equation}
solely depends on $(t, w, y)$ and solves the HJB equation
\begin{align}
    \label{EQUATION:HJB_1}
    V_{t} + \mathcal{H}(t, w, y, \nabla V, \nabla^{2} V) &= 0 &\text{for } (t, y) &\in (0, T) \times (0, \infty), \\
    \label{EQUATION:HJB_2}
    V(T, y) &= u(y) &\text{for } (t, y) &\in \partial_{P} \big((0, T) \times (0, \infty)\big)
\end{align}
with the usual elliptic HJB operator $\mathcal{H}$ and the parabolic boundary operator $\partial_{P}$. It is further known that the optimal Merton ratio in Equation~\eqref{EQUATION:OPTIMAL_CONTROL_ZARIPHOPOLOU} is independent of wealth $w$ and can only vary with $(t, y)$.

Under the logarithmic utility $u(w) = \ln(w)$, the myopic weight
\begin{equation}
    \label{EQUATION:MIOPIC_WEIGHT}
    \pi^{\ast}_{\mathrm{Heston}}(y) = (\mu - r)/y
\end{equation}
is known to be optimal~\citep{BoMu2016}. Analytic expressions for general power and exponential utilities are also provided in the latter paper, though they can lead to complex solutions making them impractical.

\section{Feedback Control via Feed-Forward ANN}
\label{SECTION:FEEDBACK_CONTROL}

Instead of solving the HJB equation to compute the optimal weight, a simple yet rigorous alternative is to impose appropriate regularity assumptions. These assumptions ensure that $V$ and, therefore, $\pi^{\ast}$ are sufficiently regular functions of their arguments. In such a case, the usual ANN approximation results hold~\citep{Ho1991}, implying that the unknown feedback control $\pi^{\ast}$ can be learned with DL over any compact set in $(t, y)$ without having to compute the value function $V$.

For any given feedback control $\pi = \pi(t, y)$, $t \in [0, T]$ and $y > 0$, let $W \equiv W(t; \pi)$, $t \in [0, T]$, denote the wealth process given by Equations~\eqref{EQUATION:MARKET_MODEL_DISCRETIZED_1}--\eqref{EQUATION:WEALTH_EVOLUTION_DISCRETE}. Thus, the expected utility in Equation~\eqref{EQUATION:UTILITY_FUNCTIONAL} can be expressed as
\begin{equation}
    \label{EQUATION:OBJECTIVE_FUNCTIONAL_DISCRETE}
    J(\pi) = \mathbb{E}\big[U\big(W_{T}^{\pi})\big)\big]
\end{equation}
assuming the initial values are fixed (viz.~Equations~\eqref{EQUATION:MARKET_MODEL_DISCRETIZED_1}--\eqref{EQUATION:WEALTH_EVOLUTION_DISCRETE}).

Instead of solving the functional optimization problem, the feedback control is represented via feed-forward ANN~\cite[Chapter~4]{Hay2009}
\begin{equation}
    \label{EQUATION:FEEDBACK_CONTROL_ANN_ANSATZ}
    \pi_{t}(y) = \mathcal{N}(t, y|\boldsymbol{\theta})
    \quad\text{ with }\quad \boldsymbol{\theta} = \big((\boldsymbol{b}_{i}, \boldsymbol{W}_{i})\big)_{i = 1, \dots, d}
\end{equation}
consisting of $d \geq 2$ layers (one input, one output and $(d - 2)$ hidden layers).
In general, an ANN can be written as a $d$-fold composition%
\footnote{``$\bigodot$'' denotes the composition operator.}
\begin{equation}
    \label{EQUATION:GENERAL_FEEDFORWARD_ANN}
    \mathcal{N}(\boldsymbol{x}|\boldsymbol{\theta}) = \Big(\bigodot_{i = 1}^{d} a_{i}\big(\boldsymbol{W}_{i} \cdot + \boldsymbol{b}_{i}\big)\Big)(\boldsymbol{x})
\end{equation}
with bias vectors $\boldsymbol{b}_{i} \in \mathbb{R}^{n_{i}}$, weight matrices  $\boldsymbol{W}_{i} \in \mathbb{R}^{n_{i} \times n_{i - 1}}$ and activation functions $a_{i} \colon \mathbb{R} \to \mathbb{R}$, $i = 1, \dots, d$.
In our specific case, $n_{0} = 2$ (two inputs), $n_{d} = 1$ (one output). As for activation functions, $a_{d}(x) \equiv x$ is chosen to allow for borrowing and short-selling, while the remaining activation functions are selected as sigmoid linear units%
\footnote{
SiLU: $a_{i}(x) = \frac{x}{1 + e^{-x}}$ for $x \in \mathbb{R}$, $i = 1, \dots, d - 1$.
} (SiLU) to ensure universality.

With this ANN ansatz, the expected utility reduces to a function
\begin{equation}
    \label{EQUATION:OBJECTIVE_VS_THETA}
    J(\boldsymbol{\theta}) = \mathbb{E}\big[U\big(W_{T}^{\pi = \mathcal{N}(\cdot|\boldsymbol{\theta}))}\big]
\end{equation}
of $\boldsymbol{\theta} \in \Theta$ over a domain in the parameter space $\Theta := \bigotimes_{i = 1}^{d} \mathbb{R}^{n_{i} \times n_{i - 1}} \times \mathbb{R}^{n_{i}}$.

Employing the Euler-Maruyama discretization (see~\ref{SECTION:DISCRETIZATION_APPENDIX}) to replace $W_{t}$ with the discrete process $W_{t}^{\Delta t}$, we finally arrive at the discrete empirical form of Equation~\eqref{EQUATION:OBJECTIVE_VS_THETA}:
\begin{equation}
    \label{EQUATION:EMPIRICAL_AVERAGE_UTILITY}
    J(\boldsymbol{\theta}) = \frac{1}{B} \sum_{b = 1}^{B} U\big(W_{T}^{\Delta t, \pi = \mathcal{N}(\cdot|\boldsymbol{\theta})), b}\big)
\end{equation}
averaged over a ``minibatch'' of $B$ solutions to discrete Equations~\eqref{EQUATION:MARKET_MODEL_DISCRETIZED_1}--\eqref{EQUATION:WEALTH_EVOLUTION_DISCRETE} computed along $B$ independent Wiener increments $(\Delta B^{S, b}_{t}, \Delta B^{Y, b}_{t})$, $b = 1, \dots, B$. With $-J(\boldsymbol{\theta})$ as a loss function, the ANN can now be effectively trained with an SGD method of choice. To avoid overtraining, minibatches were re-simulated at each step. 
If a market simulation is problematic due to speed considerations or limited data availability, sampling from a sufficiently large pre-computed batch offers a feasible alternative. However, utilization of data splitting into training, test and validation segments is crucial to protect against overtraining.

\section{Examples}
\label{SECTION:EXAMPLES}

We present two application examples to showcase how the DL approach from Section~\ref{SECTION:FEEDBACK_CONTROL} can be applied to optimal portfolio problem~\eqref{EQUATION:MARKET_MODEL_1}--\eqref{EQUATION:ISOELASTIC_UTILITY} when the risky asset is modeled using geometric Brownian motion (Section~\ref{SECTION:GBM_EXAMPLE}) or Heston model (Section~\ref{SECTION:HESTON_EXAMPLE}). In both situations, the S\&P 500 index (GSPC ticker) was selected as risky asset, while VIX index was used to calibrate Heston volatility process. Three years' worth of adjusted daily closing prices (1/1/2021--12/31/2023) were used for calibration purposes\footnote{%
Note that the number of data points (753 here) used for model calibration does not limit our ability to simulate arbitrarily many synthetic paths to train the ANN model.} 
as detailed below. The risk-free rate was set at $5\%$ p.a. The initial wealth was set at $\$1.0$.

After calibration, Equations~\eqref{EQUATION:MARKET_MODEL_DISCRETIZED_1}--\eqref{EQUATION:MARKET_MODEL_DISCRETIZED_3} were used to simulate hourly market dynamics over a one-year time horizon ($T = 1$ year with $252 \times 8.5 = 2142$ trading hours) with step size $\Delta t = \frac{1}{2142}$, while Equation~\eqref{EQUATION:WEALTH_EVOLUTION_DISCRETE} induced the wealth dynamics. The objective in Equation~\eqref{EQUATION:OBJECTIVE_VS_THETA} was optimized using Adaptive Moment Estimation (Adam) with default settings ($\rho_{1} = 0.9$, $\rho_{2} = 0.999$ and $\delta = 10^{-8}$)~\citep{KiBa2014}. Adam SGD is widely used in ML as it applies adaptive learning rates for individual parameters based on exponentially weighted moving averages of past gradients and their squares. To speed up training, batch size and ``scaling'' $\varepsilon$ were chosen adaptively in accordance with training schedules protocoled in Sections~\ref{SECTION:GBM_EXAMPLE} and \ref{SECTION:HESTON_EXAMPLE}. Initial weights were generated as i.i.d.~$\mathcal{N}(0, 0.1^{2})$ random variates, which corresponds to allocating nearly all wealth into cash. While \cite{vanStaFoLi2023} required $2.56$ million replications, $27{,}000$ (for GBM) to $45{,}000$ (for Heston) gradient evaluations were sufficient in our simulations.

All algorithm were implemented in plain \texttt{R} code and run in \texttt{RStudio} under 64-bit Windows 11 Pro OS on a Dell Precision 3581 laptop with 13th Gen Intel(R) Core(TM) i7-13800H 2.50 GHz CPU and 32 GB RAM. No third-party DL libraries or GPU acceleration were employed.

\subsection{Geometric Brownian Motion}
\label{SECTION:GBM_EXAMPLE}

Assuming a constant volatility dynamics $Y_{t} \equiv \sigma^{2}$,  Equations~\eqref{EQUATION:MARKET_MODEL_1}--\eqref{EQUATION:MARKET_MODEL_3} reduce to the GBM model
\begin{align}
    \label{EQUATION:GBM_1}
    \mathrm{d}P_{t} &= r P_{t}, & P_{0} &= 1, \\
    \label{EQUATION:GBM_2}
    \mathrm{d}S_{t} &= \mu S_{t} \mathrm{d}t + \sigma S_{t} \mathrm{d}B_{t}, & S_{0} &= S^{0}.
\end{align}
and are calibrated with usual maximum likelihood estimation~\citep{AnPo2023}.
Table~\ref{TABLE:GBM_MODEL_PARAMETERS} lists the parameter values used in simulations.

\begin{table}[h!]
    \centering
    \begin{tabular}{cc|ccc}
        \hline
        $S^{0}$ & $W^{0}$ & $r$ & $\mu$ & $\sigma$ \\
        \hline\hline
        $\$4770$ & $\$1.0$ & $0.050$ & $0.085$ & $0.176$ \\
        \hline
    \end{tabular}
    \caption{Initial GSPC price, initial wealth and estimated annualized parameter values.}
    \label{TABLE:GBM_MODEL_PARAMETERS}
\end{table}

\begin{figure}[h!]
    \centering
    \begin{subfigure}[b]{0.49\textwidth}
        \centering
        \includegraphics[width = \textwidth]{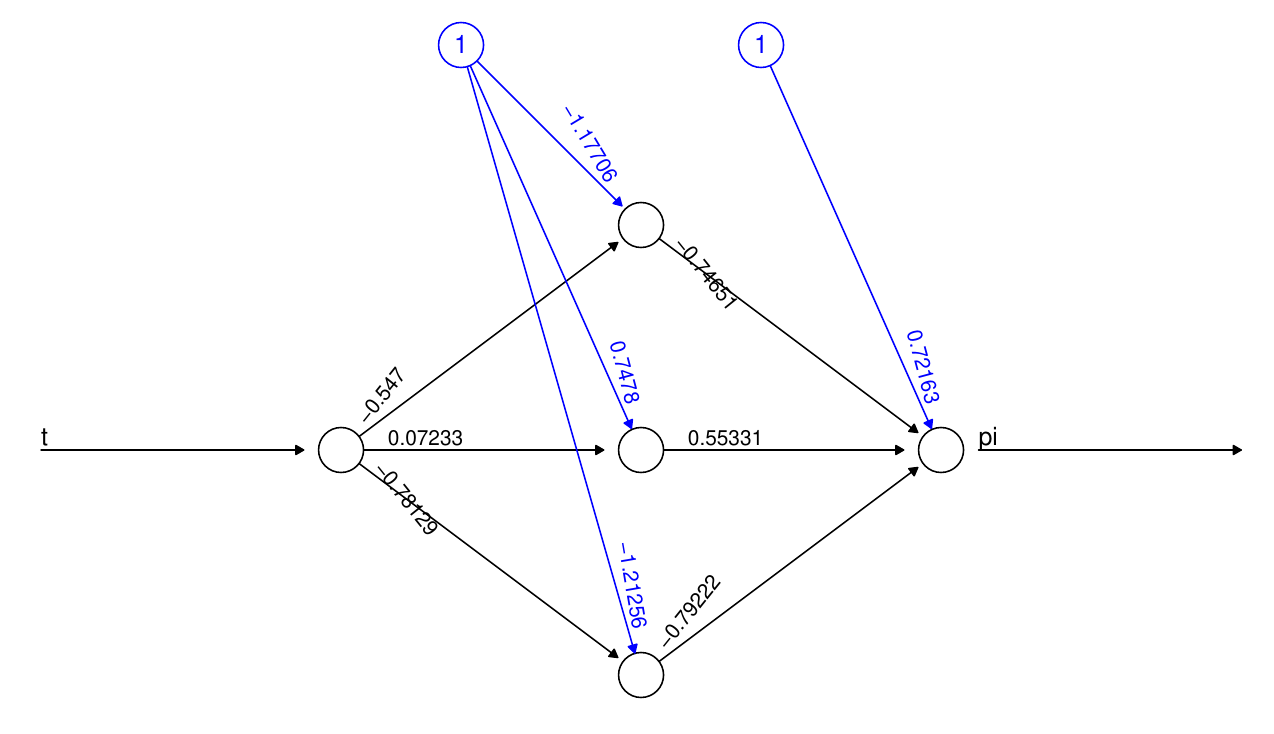}
        \caption{Trained ANN model}
    \end{subfigure}
    \hfill
    \begin{subfigure}[b]{0.49\textwidth}
        \centering
        \includegraphics[width = \textwidth]{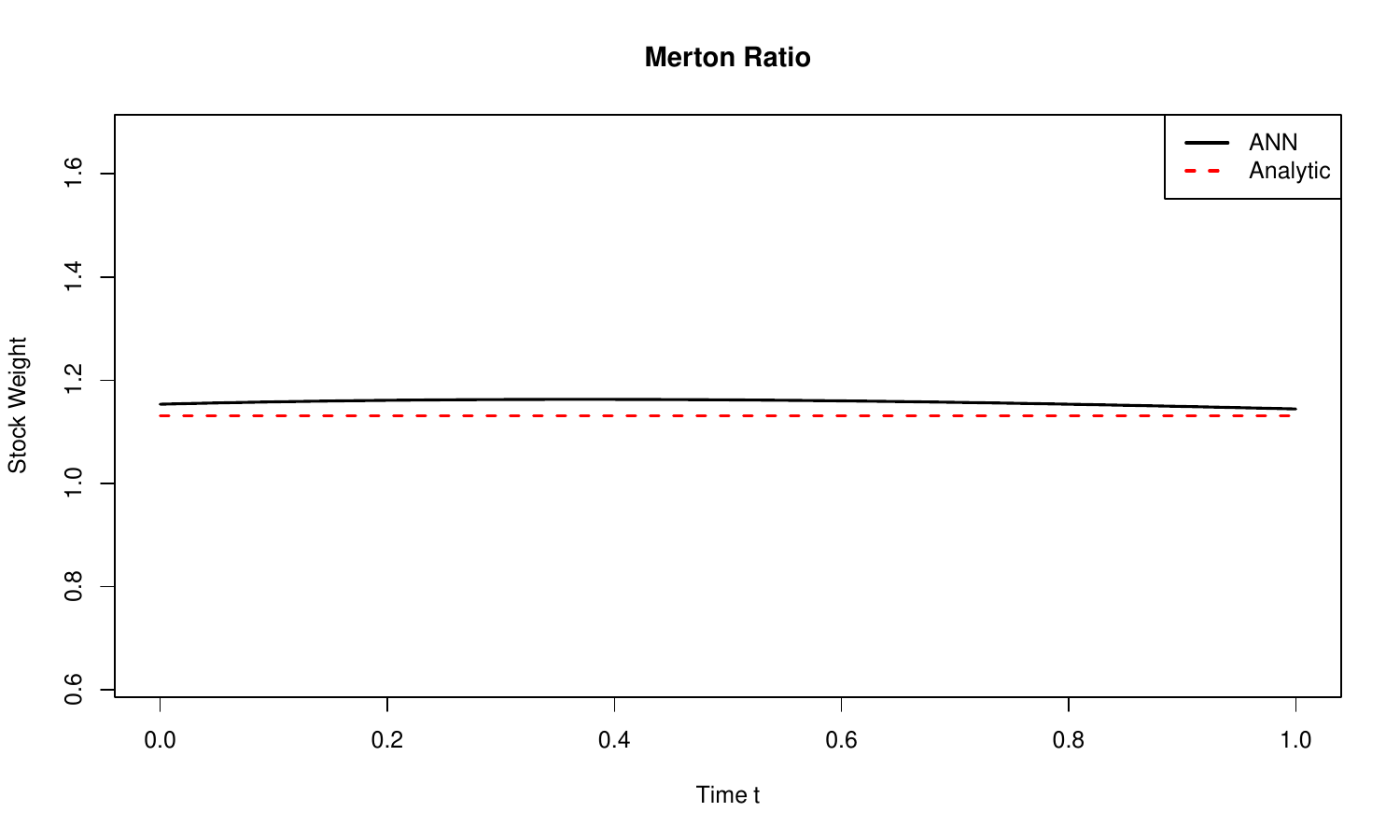}
        \caption{Estimated Merton ratio vs time $t$}
    \end{subfigure}
    \caption{Empirical results for logarithmic utility.}
    \label{FIGURE:GBM_MERTON_WEIGHT_ETA_EQUAL_ONE}
\end{figure}

Under general isoelastic utility (viz.~Equation~\eqref{EQUATION:ISOELASTIC_UTILITY}), the optimal stock weight, known as classical Merton ratio, is given by
\begin{equation}
    \label{EQUATION:CLASSICAL_MERTON_RATIO}
    \pi^{\ast}_{\mathrm{GBM}} = \frac{\mu - r}{\eta \sigma^{2}}.
\end{equation}
For comparison purposes, we chose a grid of seven equispaced $1/\eta$ values on $[0.25, 1]$. Selecting ANN with a single three-neuron hidden layer (see~ cf.~Figure~\ref{FIGURE:GBM_MERTON_WEIGHT_ETA_EQUAL_ONE}(a)), the Adam method was used to maximize the empirical power utility (viz.~Equation~\eqref{EQUATION:EMPIRICAL_AVERAGE_UTILITY}) for each $\eta$ value using the following training schedule: 100 steps with minibatch size $B = 10$ and $\varepsilon = 0.1$, 100 steps with $B = 10$, $\varepsilon = 0.05$ and 500 steps with $B = 50$, $\varepsilon = 0.01$. The training time did not exceed 3.5 hrs for each $\eta$.

\begin{figure}[h!]
    \centering
    \includegraphics[width = 0.6\textwidth]{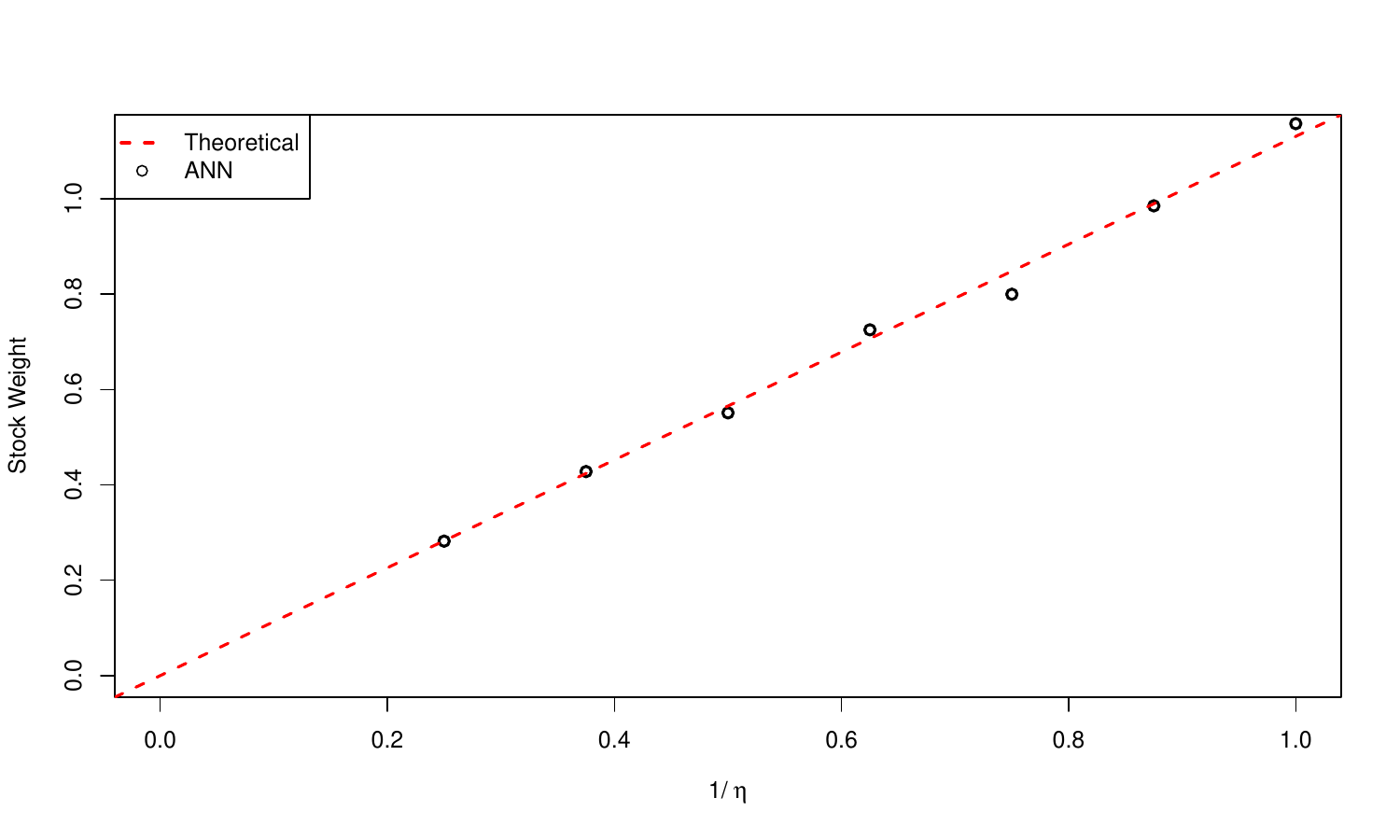}
    \caption{Merton ratio vs $1/\eta$.}
    \label{FIGURE:GBM_MERTON_WEIGHT_VS_ONE_OVER_ETA}
\end{figure}

Figure~\ref{FIGURE:GBM_MERTON_WEIGHT_ETA_EQUAL_ONE} displays the trained ANN (left panel) and the resulting weight profile $\pi(t)$ (right panel) for $\eta = 1$. The profile appears to be nearly flat closely following analytic one. Similar results were obtained for the remaining six $\eta$ values. Figure~\ref{FIGURE:GBM_MERTON_WEIGHT_VS_ONE_OVER_ETA} plots the stock weight (averaged over 500 equispaced time points) vs $1/\eta$. Good agreement with the analytic optimum can be observed. Using the discrete dynamics in Equations~\eqref{EQUATION:MARKET_MODEL_DISCRETIZED_1}--\eqref{EQUATION:WEALTH_EVOLUTION_DISCRETE}, the expected utility under both ANN and analytic optimal weights was simulated based on $10{,}000$ Monte Carlo replications as reported in Table~\ref{TABLE:GBM_UTILITY_ESTIMATES}. Again, the expected utility under ANN was on par with that under the analytic weight from Equation~\eqref{EQUATION:CLASSICAL_MERTON_RATIO} over the range of $\eta$'s considered.
\begingroup
\renewcommand*{\arraystretch}{1.4}
\begin{table}[h!]
    \centering
    \scriptsize
    \begin{tabular}{c|c|ccccccc}
        \hline
        & $1/\eta$ & 0.250 & 0.375 & 0.500 & 0.625 & 0.750 & 0.875 & 1.000 \\
        \hline\hline
        \parbox[t]{2mm}{\multirow{2}{*}{\rotatebox[origin=c]{90}{ANN}}} & Mean &
        0.05055 & 0.05563 & 0.05897 & 0.06050 & 0.06479 & 0.06643 & 0.07200 \\
        & Std error & $1.83 {\cdot} 10^{-5}$ & $4.86 {\cdot} 10^{-5}$ & $8.61 {\cdot} 10^{-5}$ & 0.00015 & 0.00019 & 0.00029 & 0.00041 \\
        \hline
        \parbox[t]{2mm}{\multirow{2}{*}{\rotatebox[origin=c]{90}{\tiny Analyt.}}} & Mean & 
        0.05027 & 0.05555 & 0.05765 & 0.06408 & 0.06387 & 0.06753 & 0.07000 \\
        & Std error & $1.77 {\cdot} 10^{-5}$ & $4.65 {\cdot} 10^{-5}$ & $9.01 {\cdot} 10^{-5}$ & 0.00014 & 0.00021 & 0.00030 & 0.00040 \\
        \hline
    \end{tabular}
    \caption{Empirically estimated (based on $10,000$ Monte Carlo replications) mean terminal utility and standard error thereof under analytic and ANN weights for various $1/\eta$ values.}
    \label{TABLE:GBM_UTILITY_ESTIMATES}
\end{table}
\endgroup

\subsection{Heston Model}
\label{SECTION:HESTON_EXAMPLE}

\begin{figure}[h!]
    \centering
    \includegraphics[width = 0.75\textwidth]{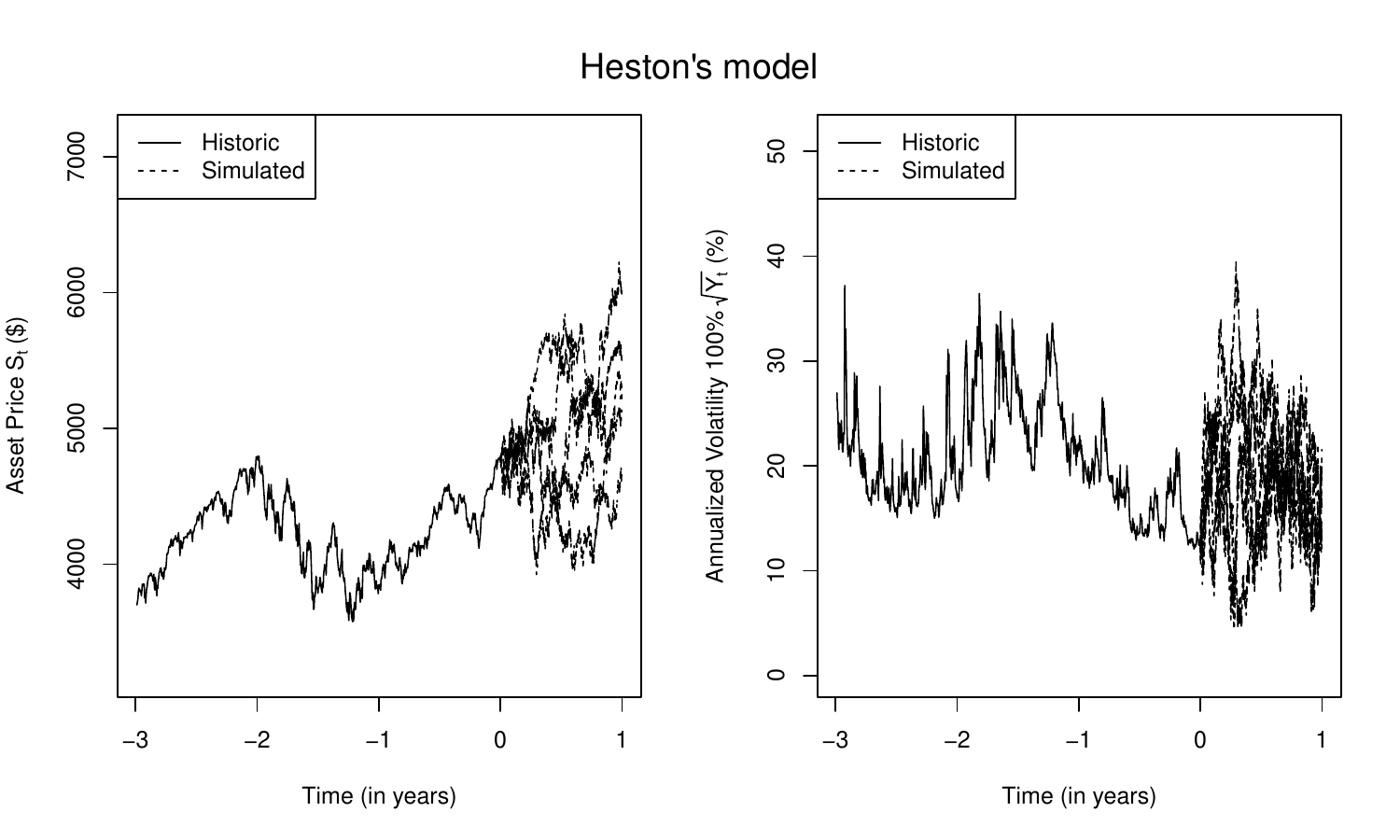}
    \caption{3 yrs of historical data and 1-yr Heston forecast for GSPC (left) and VIX (right).}
    \label{FIGURE:MARKET_DATA}
\end{figure}

Using GSPC data for $S_{t}$ and VIX for $Y_{t}$, weighted least squares were used for calibration. As before, the risk-free rate was set at $5\%$~p.a. Initial values and annualized parameters used in simulations are reported in Table~\ref{TABLE:HESTON_MODEL_PARAMETERS}. 
\begin{table}[h!]
    \centering
    \begin{tabular}{ccc|cccccc}
        \hline
        $S^{0}$ & $Y^{0}$ & $W^{0}$ & $r$ & $\mu$ & $\theta$ & $\kappa$ & $\sigma$ & $\rho$ \\
        \hline\hline
        $\$4770$ & 0.0155 & $\$1.0$ & 0.050 & 0.089 & 0.0438 & 10.5 & 0.564 & -0.712 \\
        \hline
    \end{tabular}
    \caption{Annualized estimated parameter values of the Heston model.}
    \label{TABLE:HESTON_MODEL_PARAMETERS}
\end{table}
Figure~\ref{FIGURE:MARKET_DATA} displays three years of calibration data (years -3 to 0) and five simulated one-year market paths (years 0 to 1) for risky asset $S_{t}$ (left panel) and squared volatility $Y_{t}$ (right panel). 
Unlike GBM where analytic optimal weight is known for arbitrary $\eta$, a real-valued optimal weight, being the myopic weight from Equation~\eqref{EQUATION:MIOPIC_WEIGHT}, is only known for logarithmic utility ($\eta = 1$) for Heston model~\citep{BoMu2016} limiting our comparisons to this situation.

\begin{figure}[h!]
    \centering
    \begin{subfigure}[b]{0.49\textwidth}
        \centering
        \includegraphics[width = \textwidth]{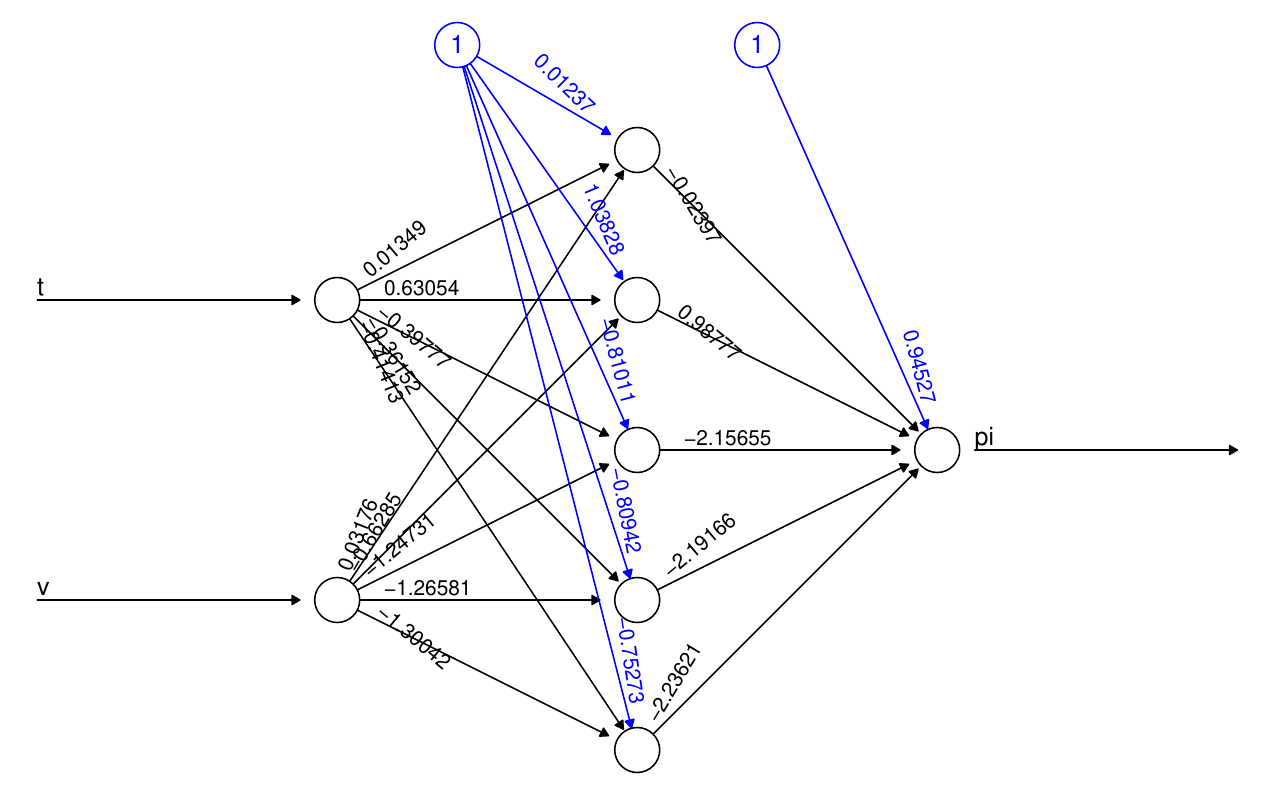}
        \caption{Trained ANN model}
    \end{subfigure}
    \hfill
    \begin{subfigure}[b]{0.49\textwidth}
        \centering
        \includegraphics[width = \textwidth]{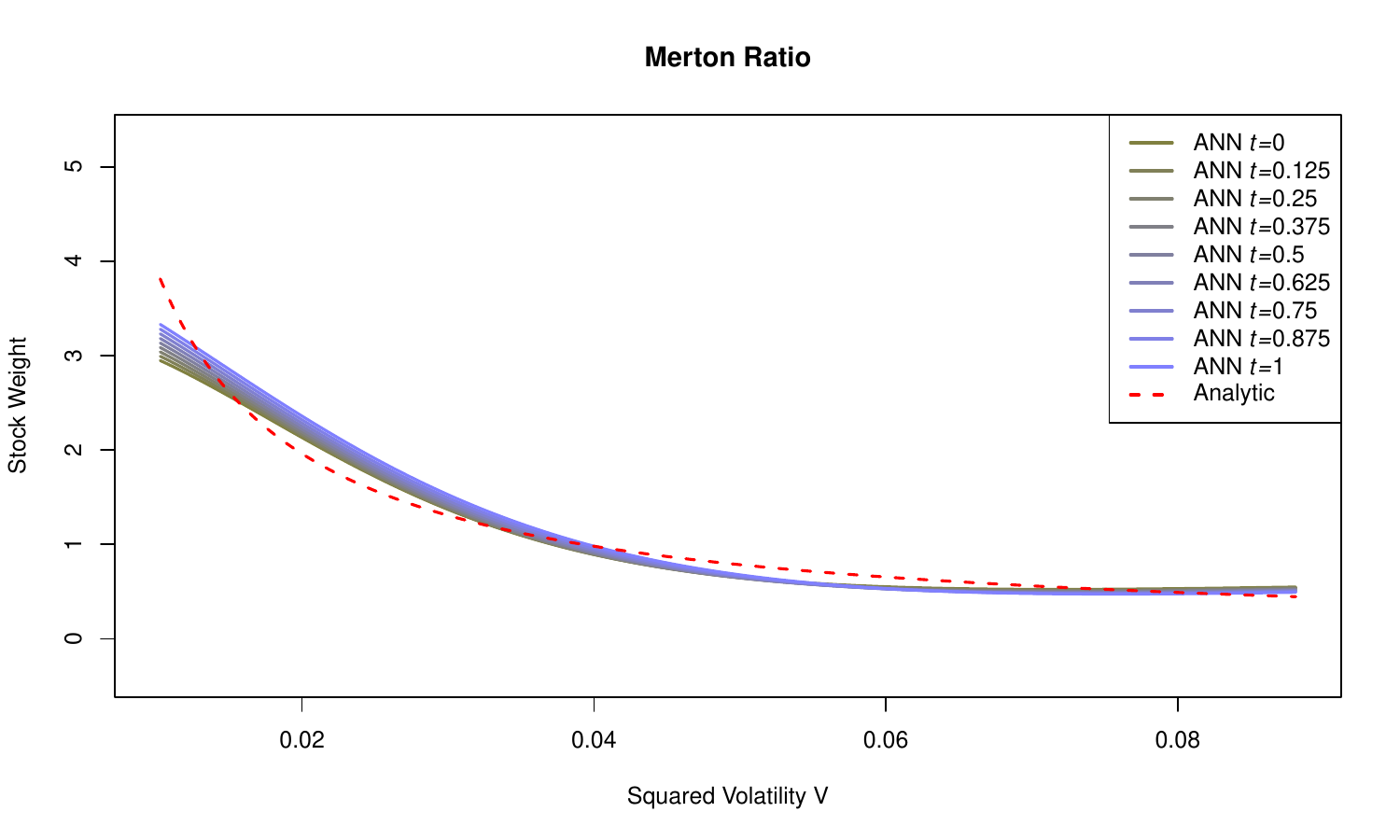}
        \caption{Estimated Merton ratio vs sqr'd volatility $y$}
    \end{subfigure}
    \caption{Results for logarithmic utility.}
    \label{FIGURE:HESTON_MERTON_WEIGHT_ETA_EQUAL_ONE}
\end{figure}

An ANN with a single five-neuron hidden layer (viz.~Figure~\ref{FIGURE:HESTON_MERTON_WEIGHT_ETA_EQUAL_ONE}(a)) was selected and trained with Adam  to maximize the empirical power utility using the following training schedule: 1500 steps with minibatch size $B = 10$ and $\varepsilon = 0.05$, 500 steps with $B = 10$, $\varepsilon = 0.01$ and 500 steps with $B = 50$, $\varepsilon = 0.01$. The training time was less than 6 hrs.

\begin{table}[h!]
    \centering
    \begin{tabular}{c|cc}
        \hline
                  & ANN & Analytic \\
        \hline\hline
        Mean      & 0.07840 & 0.07748 \\
        Std error & 0.00055 & 0.00060 \\
        \hline
    \end{tabular}
    \caption{Empirically estimated (based on $10,000$ Monte Carlo replications) mean terminal utility and standard error thereof under analytic and ANN weights for $\eta = 1$.}
    \label{TABLE:HESTON_UTILITY_ESTIMATES}
\end{table}

\begin{figure}[h!]
    \centering
    \includegraphics[width = 0.75\textwidth]{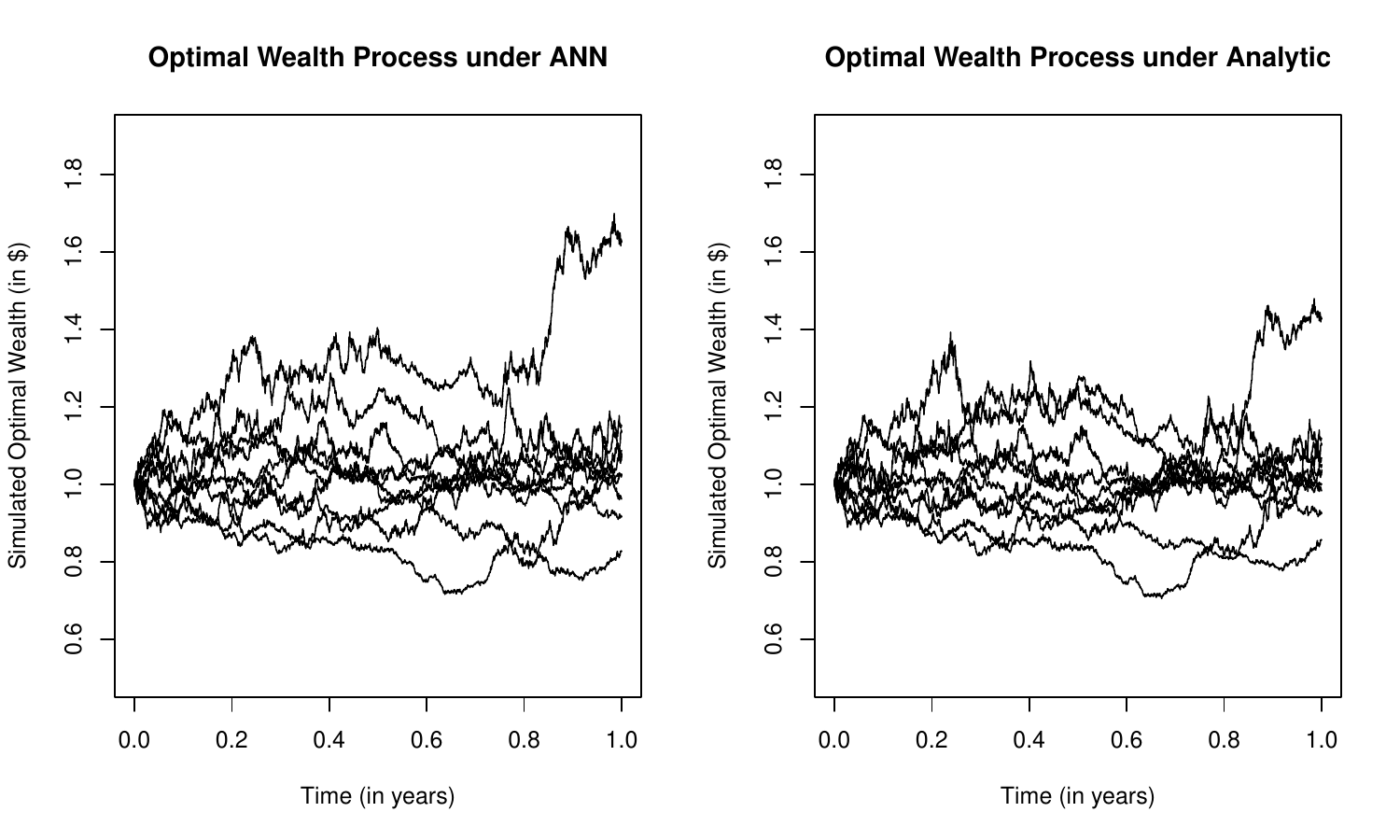}
    \caption{Simulated optimal wealth for Heston model with $\eta = 1$.}
    \label{FIGURE:OPTIMAL_WEALTH_HESTON_ETA_1}
\end{figure}

Figure~\ref{FIGURE:HESTON_MERTON_WEIGHT_ETA_EQUAL_ONE}(b) plots the optimal stock weight vs $y$ for multiple $t$ values. Over the typical range of squared volatility values (average pathwise $95\%$ range), ANN-based and analytic weights appear in good agreement. While the discrepancy grows for smaller $y$ values, such occurrences are less likely under the stochastic volatility dynamics ($Y_{t}$ long-term mean of $\theta = 0.0438$) and, thus, have less impact on the wealth process. While the analytic weight $\pi^{\ast}_{\mathrm{Heston}}$ solely depends on $y$, the empirical ANN-based weight exhibits some spurious $t$-dependence. Again, since $Y_{t}$ tend to cluster around $\theta$ for larger $t$, this discrepancy has little bearing on the utility objective under the discretized dynamics as Table~\ref{TABLE:HESTON_UTILITY_ESTIMATES} confirms. Simulated optimal wealth dynamics plotted in Figure~\ref{FIGURE:OPTIMAL_WEALTH_HESTON_ETA_1} also confirms the ANN strategy is on par with the analytic one.

\section{Summary and Conclusions}
\label{SECTION:SUMMARY_AND_CONCLUSIONS}

The empirical risk minimization approach was applied to obtain an ANN-based solution to the optimal portfolio problem under GBM and Heston models without deriving or solving the HJB equation. Calibrating the market on historic S\&P 500 and VIX data, our ANN solution was compared to the analytic one over a number of scenarios. With only 3-5 hidden neurons and a single hidden layer, the ANN results appeared to be consistent with the analytic ones both in terms of portfolio weights and expected utility values. The methodology is implemented as a standalone \texttt{R} module and can be readily applied to other stochastic volatility models, including GARCH(1, 1)~\citep{Du1997}, \cite{Kra2005} and 3/2 models~\citep{Dri2012}, etc. The approach can also be easily extended to multiple assets and/or asset classes, exotic market dynamics and many other practically relevant considerations.

\addcontentsline{toc}{section}{Supplementary Materials}
\begin{center}
	{\bf \large \uppercase{Supplementary Materials}}
\end{center}
All data and codes reported in this paper are available at: \\
\url{https://github.com/mpokojovy/StochVolPortfolioML}

\section*{Disclosure Statement}
The authors report there are no competing interests to declare.

\section*{Data Availability Statement}
All data presented in the article are available in the Supplement.

\section*{Funding}
None.

\appendix

\section{Euler-Maruyama Discretization}
\label{SECTION:DISCRETIZATION_APPENDIX}

Consider an equispaced grid $I^{\Delta t} = \{t_{k} \,|\, t_{k} = (\Delta t) k, k = 0, \dots, n\}$ with a time step $\Delta t = T/n > 0$ for some $n \in \mathbb{N}$. The Euler-Maruyama approximation $(P^{\Delta t}_{t_{k + 1}}, S^{\Delta t}_{t_{k + 1}}, Y^{\Delta t}_{t_{k + 1}})$ to $(P_{t}, S_{t}, Y_{t})$ from Equations~\eqref{EQUATION:MARKET_MODEL_1}--\eqref{EQUATION:MARKET_MODEL_3} follows the discrete dynamics
\begin{align}
    \label{EQUATION:MARKET_MODEL_DISCRETIZED_1}
    P^{\Delta t}_{t_{k + 1}} &= \big(1 + (\Delta t) r\big) P^{\Delta t}_{t_{k}}, & P^{\Delta t}_{t_{0}} &= 1, \\
    \label{EQUATION:MARKET_MODEL_DISCRETIZED_2}
    S^{\Delta t}_{t_{k + 1}} &= \big(1 + (\Delta t) \mu\big) S^{\Delta t}_{t_{k}} + \sqrt{Y^{\Delta t}_{t_{k}}} \, S^{\Delta t}_{t_{k}} \, \Delta B^{S}_{t_{k}}, & S^{\Delta t}_{t_{0}} &= S^{0}, \\
    \label{EQUATION:MARKET_MODEL_DISCRETIZED_3}
    Y^{\Delta t}_{t_{k + 1}} &= \left(1 + (\Delta t) \kappa \big(\theta - Y^{\Delta t}_{t_{k}}\big)\right) + \sqrt{Y^{\Delta t}_{t_{k}}} \, \Delta B^{Y}_{t_{k}}, & Y^{\Delta t}_{t_{0}} &= Y^{0}
\end{align}
where $(\Delta B_{t_{k}}^{S}, \Delta B_{t_{k}}^{Y}) \stackrel{\mathrm{i.i.d.}}{\sim} \mathcal{N}(\boldsymbol{0}, (\Delta t) \boldsymbol{\Sigma})$ for $k \geq 0$ with $\boldsymbol{\Sigma} = \begin{pmatrix}
    \rho & 1 \\
    1 & \rho
\end{pmatrix}$, while the discrete wealth equation is given by
\begin{align}
    \label{EQUATION:WEALTH_EVOLUTION_DISCRETE}
    \begin{split}
        W^{\Delta t}_{t_{k + 1}} &= W^{\Delta t}_{t_{k}} \Big(1 + (\Delta t) (1 - \pi^{\Delta t}_{t}) \tfrac{P^{\Delta t}_{t_{k + 1}} - P_{t_{k}}}{P^{\Delta t}_{t_{k}}} + (\Delta t) \pi^{\Delta t}_{t} \tfrac{S^{\Delta t}_{t_{k + 1}} - S_{t_{k}}}{S^{\Delta t}_{t_{k}}}\Big), \\
        W^{\Delta t}_{t_{0}} &= W^{0}
    \end{split}
\end{align}
with the discretized stock weight process $\pi^{\Delta t}_{t}$.

\end{document}